# Cool windows: simultaneously engineering high visible transparency and strong solar rejection

*Yeonghoon Jin,*[1,2,#] *Seungwon Kim,*[1,#] *Tanuj Kumar,*[2] *Mikhail A. Kats,*[2*] *and Kyoungsik Yu*[1,*]

[1]School of Electrical Engineering, Korea Advanced Institute of Science and Technology (KAIST), Daejeon 34141, Republic of Korea

[2]Department of Electrical and Computer Engineering, University of Wisconsin-Madison, Madison, WI 53706, USA

[#]These authors contributed equally

E-mail: mkats@wisc.edu, ksyu@kaist.edu

## Abstract

For window applications in hot climates, it is desirable to have windows with high visible transparency, while maintaining strong reflectance in both the ultraviolet and near infrared, to minimize unwanted heat gain. Given that more than 70% of incident solar energy is at wavelengths shorter than 1000 nm, achieving spectrally abrupt transitions from transparent to reflecting at the boundaries of the visible is essential. Such abrupt transitions at multiple wavelengths typically would require tens of dielectric layers, which is impractical for most window applications. Here, we propose and realize a structure comprising only eight planar layers that achieves sharp reflectance changes at ~390 and ~680 nm, resulting in high visible transmittance (>70%), and high near-infrared (>80%) and UV (>60%) reflectance, as well as high mid-infrared emissivity (>90%) for additional radiative cooling. We demonstrate that our engineered window reduces air temperature by up to 3.8 °C inside an enclosed box simulating a vehicle, compared to a box with a reference window.

## Introduction

The inside of a car parked in sunlight on a hot summer day can quickly reach lethal temperatures, for example reaching 70 °C when the outside temperature is 40–45 °C.[1] This is because the windows form a greenhouse, passing most solar wavelengths while rejecting the mid-infrared thermal radiation that would otherwise enable cooling. The solar spectrum reaching the Earth's surface has meaningful power for wavelengths from 0.3 to 2.5 μm, and nearly all but the shortest of these wavelengths are transmitted through typical windows, resulting in the temperature rise inside of buildings or vehicles.

In this paper, we aim to design a highly visible-transparent window for hot-weather operation. In particular, we aim to have sufficiently high transparency across most of the visible (400–680 nm) range, along with high reflectance in the ultraviolet (300–400 nm) and the near-infrared (680–2500 nm) to prevent unwanted heating (**Figure 1b**). Notably, the highest near-infrared and near-UV intensities of sunlight are in the wavelength ranges just outside the visible, and therefore an abrupt change in optical reflectance at the visible boundaries (both 400 and 680 nm) is necessary to maximize heat rejection (**Figure 1b**).

Along with the solar spectrum management, passive radiative cooling via mid-infrared thermal emission from a window toward the sky also contributes to net heat transfer.[2-3] Although solar management[4-7] and passive radiative cooling[2-3] have been studied independently for decades, the combination of the two strategies has emerged relatively recently.[8-13] However, numerous previous works have tended to emphasize mid-infrared thermal emission engineering, rather than focusing on the solar management. Given that the incident solar energy is around 1000 $W/m^2$ (though the specific value depends on the location, date, and time), whereas the cooling potential via thermal radiation in many practical scenarios is often less than ~150 $W/m^2$,[14] we consider solar management to be the primary factor governing the overall heat transfer, with thermal radiation making a meaningful but smaller contribution.

Simultaneous solar management and radiative cooling can be achieved by integrating solar-rejecting structures (generally visible-transparent but infrared-reflective) with materials that have high mid-infrared emissivity, such as polymers and oxides.[15] Solar management examples include dielectric-metal-dielectric (DMD) thin-film structures[4-7, 16] and conducting transparent oxide films[9, 17-18]. However, such strategies generally exhibit gradual changes in reflectance and transmittance, limiting the rejection of the unwanted portions of the solar spectrum. Both DMD structures and conducting oxide films[9, 17] generally have lower-than-ideal reflectance at short near-infrared wavelengths (*e.g.*, 700–950 nm) as shown in **Figure 2e** for DMD structures.

Another example of integrating solar management with radiative cooling is to utilize metallic nanostructures designed to resonantly reflect in the near infrared, integrated with polymers or oxides to enhance radiative cooling.[19-21] However, like DMDs and transparent conducting films, these metallic nanostructures also have insufficiently sharp transitions from transparent to reflective as a function of wavelength; furthermore, nanostructures can result in visible scattering which reduces visual clarity.

**Figure 1c** shows the solar-spectrum weighted visible transmittance $T_{vis}$ (400–680 nm) and near-infrared reflectance $R_{IR}$ (680–2500 nm) of various structures found in the literature. The DMD structures show $R_{IR}$ of 60~70%,[4-7] the conducting oxides-based structures show $R_{IR}$ less than 30%, and the metallic nanostructures show $R_{IR}$ lower than ~40%.[19-21]

There have been several attempts to demonstrate abrupt changes in reflectance and transmittance at the boundaries of the visible range. For example, abrupt changes have been realized using a 30-layer dielectric stack,[12] and via a silver/silica multilayer stack,[13] in both cases resulting in high visible transmittance ($T_{vis}$) and $R_{IR}$ (the half-filled circles in **Figure 1c**). However, none of them demonstrated $T_{vis}$ > 70% and $R_{IR}$ > 80% simultaneously; we set the standard for $T_{vis}$ at 70% because the visible transmittance for vehicle windshields must be higher than 70%, according to the U.S. National Highway Traffic Safety Administration.[22] Note that one structure gets close,[12] reporting $T_{vis}$ ~ 78% and $R_{IR}$ ~ 77%, but is composed of 30 dielectric layers, which is impractical for most window applications.

Here, we propose and demonstrate a cool window coating that satisfies high $T_{vis}$ (>70%) and $R_{IR}$ (>80%) using just 8 layers. The structure of the coating is a combination of what we refer to as a dual-band selective reflector (DBSR), a dielectric-metal-dielectric (DMD) layer that provides long-wavelength (>1000 nm) reflectance, and a transparent layer of polydimethylsiloxane (PDMS) with high mid-infrared emissivity. The DBSR is a modified distributed Bragg reflector (DBR) with a photonic bandgap in the short-wavelength near-infrared (680–950 nm), that we have engineered to have a second photonic bandgap in the ultraviolet (350–400 nm), without increasing the number of layers or the overall structure thickness compared to the conventional Bragg reflector. Our cool window shows an *air* temperature reduction inside an enclosure of as much as 3.8 °C compared to a reference piece of glass.

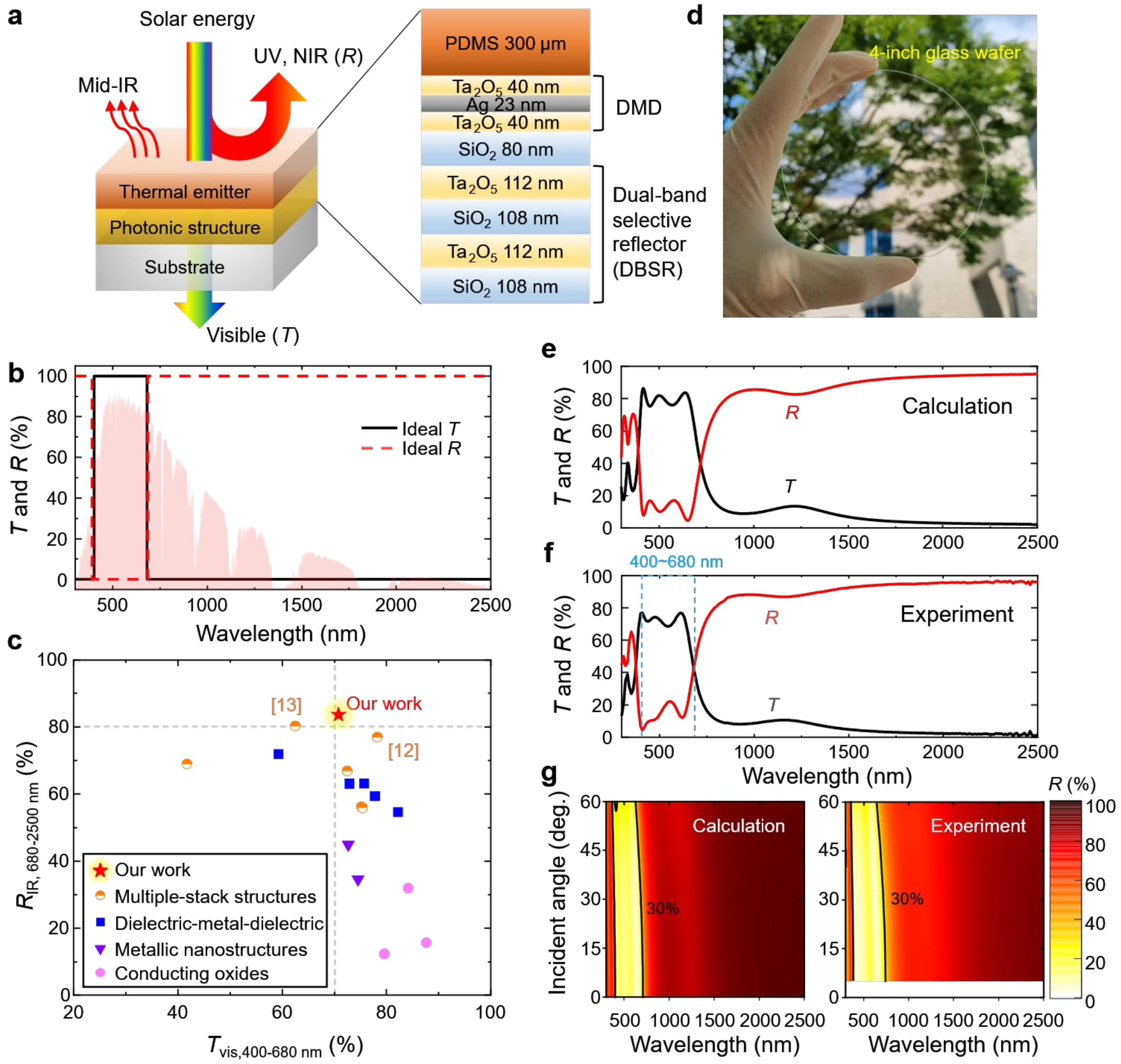

**Figure 1.** Optical design and properties of our photonic structure. **(a)** Schematic of the layered structure. **(b)** Transmittance ($T$) and reflectance ($R$) spectra of an ideal heat-rejection window. The red background is the solar spectrum (AM1.5). **(c)** Comparison of the solar-spectrum-weighted visible transmittance ($T_{vis}$, 400–680 nm) and IR reflectance ($R_{IR}$, 680–2500 nm) between our photonic structure and previous works. **(d)** A photo of our photonic structure, fabricated on a 4-inch glass, which shows its high visible transparency. **(e, f)** Calculated (e) and measured (f) $R$ and $T$ spectra. **(g)** Calculated (left) and measured (right) angle-dependent $R$ spectra. The input light was assumed to be unpolarized.

## Results

### Optical properties of our cool window

A schematic of our cool window is shown in **Figure 1a**, with a photo of a realization on a 4-inch glass wafer shown in **Figure 1d**, demonstrating high visible transparency. **Figure 1e** shows the calculated reflectance ($R$) and transmittance ($T$) spectra of the cool window without the top PDMS layer, and **Figure 1f** shows the experimentally measured spectra. The calculation and measurement show excellent agreement, which confirms that the actual thickness of each layer is comparable to the values used in the calculation. Abrupt changes in $R$ and $T$ can be observed around 390 and 680 nm, as desired for efficient heat rejection. The average transmittance in the visible (400–680 nm) range, $T_{vis}$, is ~71% and the average reflectance in the 680–2500 nm range, $R_{IR}$ is ~83% (**Figure 1c**). Note that high $T_{vis}$ and $R_{IR}$ can be achieved when the spectral transition at the visible boundaries is sufficiently sharp. The angle-dependent reflectance and transmittance spectra do not change appreciably as a function of incidence angle, at least up to 60° for unpolarized light (**Figure 1g**).

**Optical design of our cool window**

The solar management of our cool window is achieved via two components: the top dielectric-metal-dielectric (DMD) and the bottom dual-band selective reflector (DBSR). The top DMD ($Ta_2O_5$ 40 nm/Ag 23 nm/$Ta_2O_5$ 40 nm) is intended to reflect light at wavelengths longer than 1000 nm, but it lacks high reflectance in the wavelength ranges of 350–400 nm and 680–1000 nm (**Figure 2e**, the black dashed line). To address this limitation, we used our DBSR concept to create two photonic bandgaps on both sides of the visible spectrum. The DBSRs are implemented using pairs of $SiO_2$ (108 nm)/$Ta_2O_5$ (112 nm) films, where the thicknesses are slightly different than those of a conventional DBR, as described below. $SiO_2$ and $Ta_2O_5$ are standard thin-film optics materials, and were chosen because of their significant refractive index difference (the refractive index of $SiO_2$ is ~1.45 and that of $Ta_2O_5$ is ~2.1 in the visible to near infrared). The large difference in the refractive index is required to achieve sufficient reflectance with minimal number of layers, as in conventional DBRs.

The calculated reflectance spectra of DBSRs are shown in **Figure 2a**, for increasing numbers of $SiO_2$/$Ta_2O_5$ pairs. The reflectance in the wavelength ranges of 350–400 nm and 680–950 nm increases with the number of pairs, while maintaining high visible transmittance. We would like to emphasize here that our DBSRs are different from DBRs because conventional DBRs cannot simultaneously achieve high $R$ at the UV and near-IR ranges (350–400 nm and 680–950 nm); for example, **Figure 2b** shows the calculated reflectance of conventional DBRs with a target center wavelength of 780 nm, where one pair is composed of $SiO_2$ (134 nm)/$Ta_2O_5$ (94 nm), with each thickness corresponding to $\lambda/4n$, where $\lambda$ is the wavelength of 780 nm and $n$ is the refractive index ($n_{SiO2}$ = 1.45 and $n_{Ta2O5}$ = 2.08 at 780 nm). The DBR spectra do not show high $R$ in the UV range, while their spectra in the near-infrared are almost identical to those of our DBSRs (compare **Figure 2a** vs. **Figure 2b**).

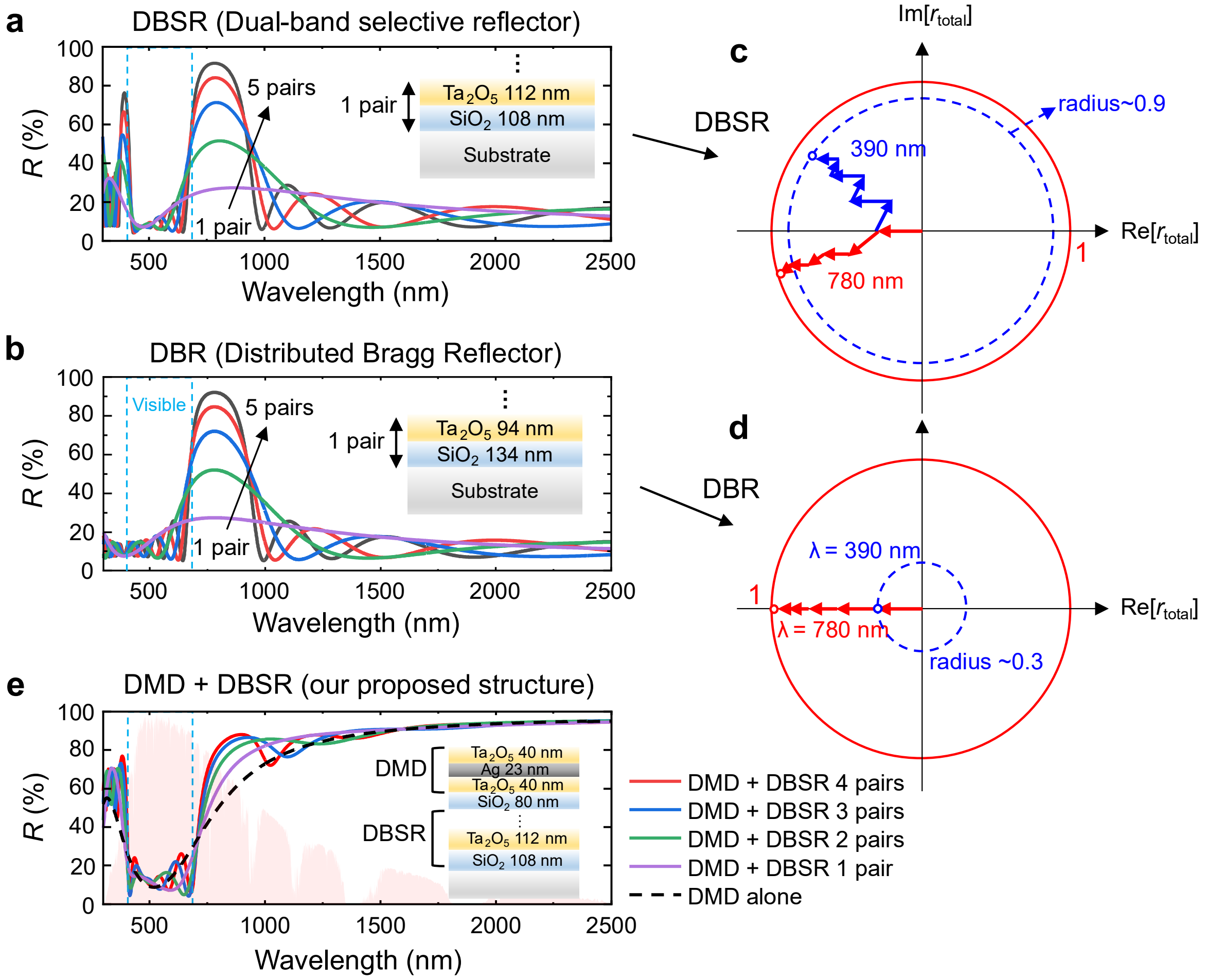

**Figure 2. Achieving simultaneous rejection of near-infrared and near-UV light**. **(a)** Calculated reflectance spectra of our dual-band selective reflectors (DBSRs), from 1 to 5 pairs. One pair corresponds to layers of $SiO_2$ (108 nm) and $Ta_2O_5$ (112 nm). **(b)** Calculated reflectance spectra of reference DBRs for the same materials and number of pairs, but where the layers are $SiO_2$ (134 nm) and $Ta_2O_5$ (94 nm), each film corresponding to a quarter wavelength at $\lambda$ = 780 nm. Unlike the DBSR, the DBR does not have a secondary reflectance peak in the near UV. **(c, d)** Visual phasor addition for partial waves contributing to the overall reflection coefficient, plotted in the complex plane for (c) our DBSR and (d) our reference DBR. The amplitude of the overall reflection coefficient, $|r_{total}|$, at the target wavelength ($\lambda$ = 780 nm) is ~1 for both cases. However, at $\lambda$ = 390 nm, only the DBSR shows high $|r_{total}|$. **(e)** Calculated reflectance spectra of the integrated photonic structure, where the dielectric-metal-dielectric (DMD) is placed on top of a DBSR with different numbers of pairs. There is an 80-nm-thick $SiO_2$ intermediate layer as an optical spacer between the DMD and the DBSR.

To briefly explain the difference between conventional DBRs and our DBSRs, the thickness of conventional DBRs ($\lambda_{target}/4n$) is designed to produce constructive interference for reflected light, which occurs at $\lambda_{target}/(2m-1)$, where $m$ is a positive integer. For example, for $\lambda_{target}$ = 780 nm, where the first high-reflectance mode ($m$ = 1) occurs, the second mode ($m$ = 2) appears at 780/3 = 260 nm. This means that it is difficult to achieve high reflectance in both the 350–400 and 680–950 nm ranges using conventional DBRs. **Figure 2d** visually shows the phasor addition of the partially reflected waves of the 5 pair DBR. The "phasor trajectory" for $\lambda_{target}$ = 780 nm (red) shows complete constructive interference with the reflection coefficient amplitude of 1, whereas that of 390 nm shows the amplitude of ~0.3 (this corresponds to a power reflectance of 0.09).

However, we found that if the thickness of each layer is made to deviate slightly from $\lambda_{target}/4n$, which results in a slight mismatch of phase between each partial wave, the wavelengths of higher-order resonances ($m = 2, 3\ldots$) can be manipulated while maintaining the first-order resonance near ~780 nm and still achieving near-unity reflectance. The reflectance spectra of DBSRs in **Figure 2a** show high $R$ at both the first mode around $\lambda = 780$ nm and at the second mode around $\lambda = 390$ nm, per our design. See **Note S1** for detailed comparison of the working principles of DBRs and DBSRs.

To exploit the advantages of both the DMD and DBSRs, we stacked the DMD structure and the DBSRs (1–4 pairs), with an 80-nm-thick $SiO_2$ intermediate layer as an optical spacer. **Figure 2e** shows the calculated reflectance spectra, respectively, of the combined structures, which show abrupt changes around both $\lambda = 390$ and 680 nm.

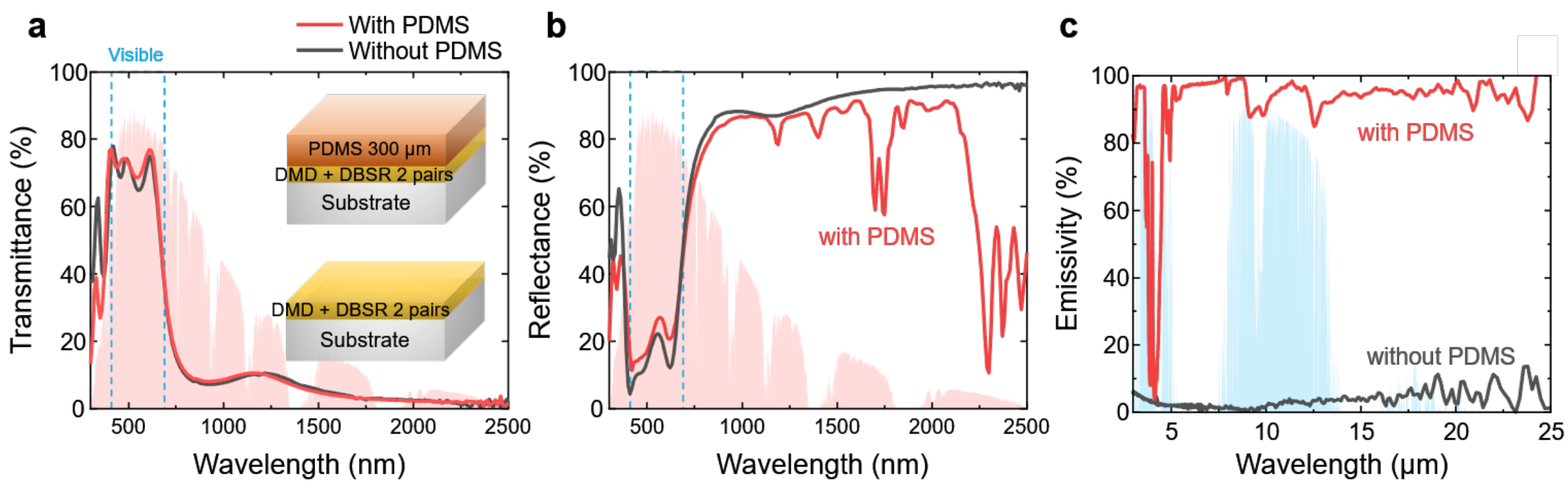

**Figure 3.** Measured optical properties with and without a thermal-emissive layer (300-μm-thick PDMS) on the DBSR/DMD stack ($Ta_2O_5$ 40 nm/Ag 23 nm/$Ta_2O_5$ 40 nm/$SiO_2$ 80 nm/$Ta_2O_5$ 112 nm/$SiO_2$ 108 nm/$Ta_2O_5$ 112 nm/SiO2 108 nm/substrate). **(a-c)** Measured **(a)** transmittance, **(b)** reflectance, and **(c)** emissivity of the DBSR/DMD stack with and without the PDMS layer.

The 2-pair DBSR/DMD stack was fabricated, and **Figure 3** shows the measured transmittance, reflectance, and emissivity spectra. Due to the presence of the Ag layer, the DBSR/DMD stack has both very low transmittance and very low emissivity in the mid-infrared (**Figure 3c**). To significantly increase the emissivity (to facilitate radiative cooling of the window), we added a 300-μm-thick PDMS layer on top because PDMS is visible-transparent, but highly emissive in the mid-infrared due to molecular vibrational resonances.[13, 23] The resulting cool window has an average emissivity ($\varepsilon$) of 94% in the 3–25 μm range, while the DBSR/DMD stack alone has an average $\varepsilon$ lower than 5% (**Figure 3c**). The PDMS layer barely affects the $T$ as shown in **Figure 3a**, but it does absorb in certain near-infrared wavelengths (**Figure 3b**). However, the portion of the solar energy in those ranges is small, so that does not significantly affect the performance of the cool window.

**Outdoor temperature experiment**

To evaluate the cooling performance of our cool window, we conducted an outdoor temperature experiment on Oct 10, 2023, in Daejeon, South Korea, during the day. We prepared four polystyrene boxes ($26 \times 26 \times 27$ $cm^3$) with 2 cm thickness and cut a square hole ($6.4 \times 6.4$ $cm^2$) in the top of each box (**Figure 4a**). We painted the interiors of the boxes black to mimic the interior of a vehicle and placed four thermocouples in each box to measure the air temperature (the thermocouples were suspended in air). The ambient temperature outside of the box was measured by a thermocouple under a polystyrene foam to avoid direct sunlight. We prepared four samples on 4-inch glass substrates: glass, glass/PDMS, glass/2-pair DBSR/DMD structure, and glass/2-pair DBSR/DMD/PDMS; in all cases, the PDMS layer was 300 μm thick. We attached each of these samples directly below the holes in the boxes. The measured $R$ and $T$ spectra of the glass and glass/PDMS samples, spanning from the ultraviolet to mid-infrared range, are shown in **Note S2**. Finally, we placed a low-density polyethylene film on top of each hole to minimize convection.[3]

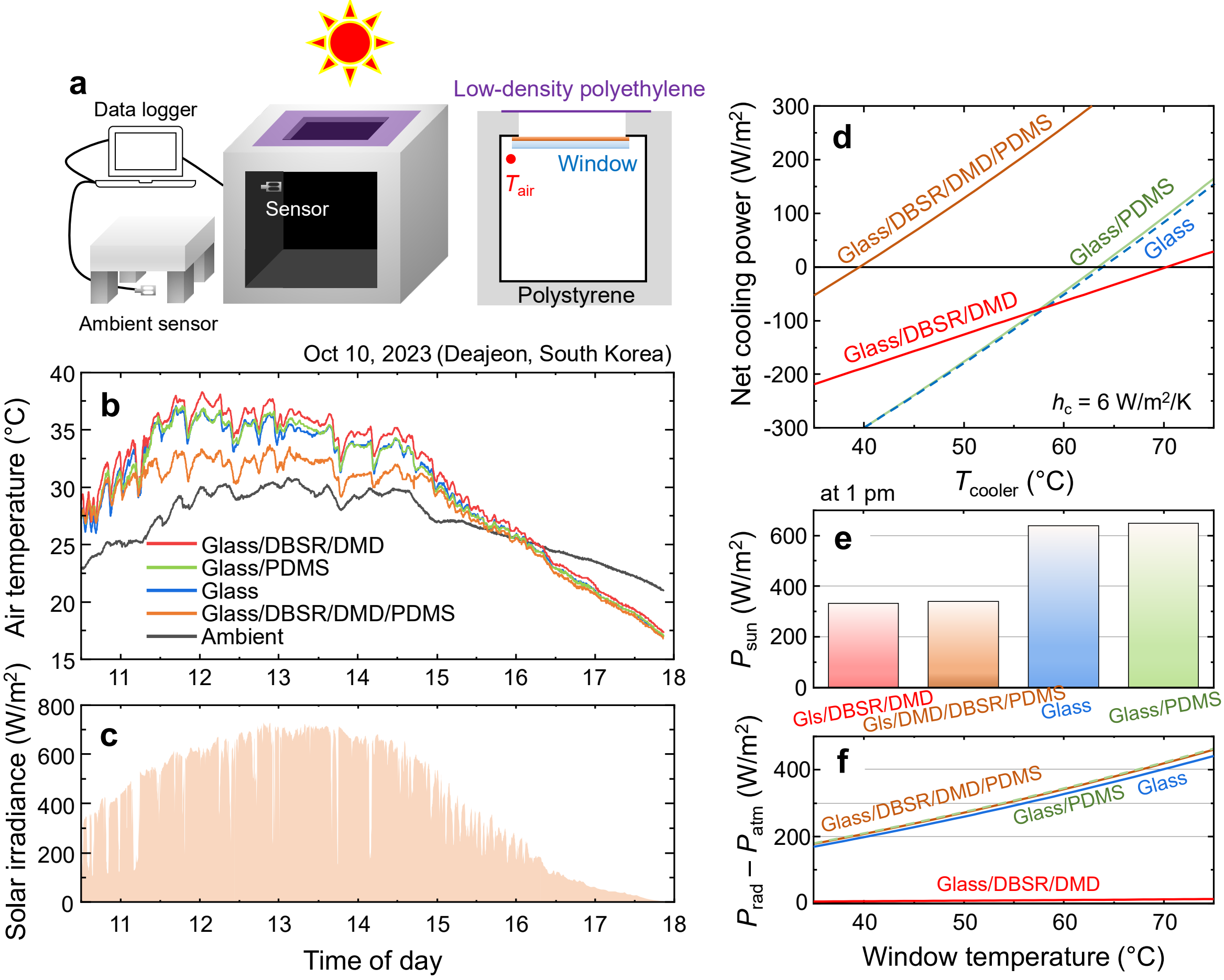

**Figure 4.** Outdoor temperature experiment. **(a)** Depiction of our outdoor experiment setup: four polystyrene boxes with a square hole on the top, painted black on the inside to mimic a vehicle. Each sample (glass, glass/PDMS, glass/DBSR (2-pair)/DMD, glass/DBSR (2-pair)/DMD/PDMS) was placed directly beneath the hole of each box, and a low-density polyethylene film was placed over the hole to minimize convection. An air temperature sensor was placed inside the box, avoiding direct sunlight, and suspended in air. An ambient temperature sensor was placed under a polystyrene foam to avoid direct sunlight. All four samples were measured simultaneously. **(b, c)** Measured air temperatures (b) and solar irradiance (c) of the four samples and the ambient temperature on October 10, 2023, at KAIST, in Daejeon, South Korea (36.27°N/127.36°E). Our cool window (glass/DBSR/DMD/PDMS) shows a temperature up to 3.8 °C lower than the glass sample, and 4.5 °C lower than the glass/DBSR/DMD without the PDMS layer. **(d)** Calculated net cooling power of the four samples as a function of the window temperature, including incident solar radiation and radiative and nonradiative cooling. **(e)** Calculated solar absorption in both the cool window, and inside the box ($P_{sun}$ defined as 715 W/m$^2$ × (1 – reflectance of the cool window)). **(f)** Calculated net radiative thermal flux, $P_{rad}$ – $P_{atm}$, where $P_{rad}$ represents the thermal radiation from the cooler and $P_{atm}$ represents the atmospheric radiation absorbed by the cool window; this was calculated using our gradient atmospheric model.[14] Note that in (f), sunlight is not included.

**Figure 4b** shows the measured air temperature inside the box for the four different windows, along with the ambient temperature, while **Figure 4c** shows the measured solar irradiance. During the daytime, the cool window (glass/DBSR/DMD/PDMS) shows the lowest temperatures, up to 3.8 °C lower than the bare glass window.

**Figure 4d** shows the calculated net cooling power of each sample as a function of the window temperature. We used our recently reported gradient atmospheric model, which considers altitude-dependent atmospheric gas composition and temperature profiles[14], to calculate the cooling power. Refer to ref.[14] and **Note S3** for details of the cooling power calculation. The equilibrium temperature, at which the net cooling power equals zero, is the lowest for the glass/DBSR/DMD/PDMS and highest for the glass/DBSR/DMD, showing good agreement with our experiment.

**Figure 4e** shows the solar absorption of the window and the black-painted box, $P_{sun}$ defined as $(1 - R) \times 715$ W/m$^2$; 715 W/m$^2$ is the solar irradiance at 1 pm. **Figure 4f** shows the net radiative thermal flux, $P_{rad} - P_{atm}$, where $P_{rad}$ represents the power emitted from the window and $P_{atm}$ represents the power emitted from the atmosphere and absorbed by the window (not accounting for sunlight). Although the solar absorption of the glass/DBSR/DMD and the glass/DBSR/DMD/PDMS samples are nearly identical (**Figure 4e**), the glass/DBSR/DMD/PDMS sample demonstrates higher performance because of its higher thermal emissivity (**Figure 4f**), resulting in a temperature drop up to 4.5 °C.

In **Figure 4b**, after approximately 4 PM, the air temperatures for all cases became lower than the ambient temperature. This is primarily due to cooling of the polystyrene box via thermal radiation, given the high emissivity of polystyrene, combined with the reduced incident solar power in the evening.

Note that in our experiments, all of the coatings were on the outside of the glass window. For this reason, the glass/DBSR/DMD sample results in higher air temperature inside, compared to the glass-only sample (**Figure 4b**), because even though the glass/DBSR/DMD film has solar transmittance ~300 W/m$^2$ lower than the glass-only sample, it also has much lower infrared emissivity. In our glass/DBSR/DMD/PDMS samples, the PDMS provides very high emissivity. However, the PDMS layer may not be necessary if the DBSR and DMD layers are coated on the inside of the glass rather than on the outside, since bare glass has relatively high emissivity; this configuration also makes it easier to protect the coatings against damage from the outside.

## Discussion

The ideal design of a window in a hot environment has slightly paradoxical needs; it needs to be transparent over the visible to provide visibility, but reflect as much sunlight as possible, despite the fact that a lot of the solar spectrum is in and around the visible range. Our proposed solution is a highly transparent window across most of the visible (400-680 nm), which has sharp transitions in the spectrum to high reflectance in the ultraviolet and near-infrared, just outside of the visible range. We accomplish these sharp transitions using just 7 optical layers by combining a metal film for near-infrared reflectance with a special modified Bragg distributed reflector which has sharp peaks in both the near-infrared and the near-ultraviolet. Our fabricated cool window has high visible transparency ($T_{vis}$ > 70%) while maintaining high reflectance in the ultraviolet ($R_{UV}$ > 60%) and the near infrared ($R_{IR}$ > 80%). The spectral features persist over a broad range of angles. We also added an additional PDMS layer on top to maximize mid-infrared emissivity and therefore maximize cooling of the window. Our cool window achieves an air temperature reduction of up to 3.8 °C compared to a reference piece of glass. Refer to **Note S4** for the energy-saving benefit of our cool window and **Note S5** for the practical design implemented to prevent the oxidation or sulfidation of Ag.

Our present, static cool window is designed for use primarily in hot environments. Windows that can adjust their optical properties are actively being explored, for example using tunable optical materials.[21, 24-26] The ideal design of such dynamic windows may have the capability to control the visible and infrared regions independently depending on the weather conditions. For example, it may be desirable to have windows switch between transparent and opaque as a function of temperature or degree of sunlight. In addition, solar near-infrared needs to be reflected and mid-infrared thermal emission needs to be high for cooling in hot environments, while near-infrared transparency and low mid-infrared emissivity may be desired for cold environments. Although there have been many significant advancements in dynamic tunability,[21, 24-26] independent control over different spectral ranges and responsiveness to temperature and solar brightness remains challenging.

We believe that even in the absence of tunability, deliberate solar management using spectrally sharp transitions between the visible and the adjacent near-UV and near-IR wavelength ranges can provide meaningful advantages in hot climates, reducing the risk of vehicular hyperthermia in parked cars and providing energy savings in buildings.

**Methods**

*Sample Fabrication and optical characterization:* The oxide materials ($SiO_2$ and $Ta_2O_5$) were deposited using an electron-beam evaporator, and the Ag layer was deposited using a thermal evaporator; both evaporators were in the same chamber. The base pressure of the chamber was ~$10^{-6}$ Torr. The deposition rate for the oxide materials was 2–3 Å/s, and ~3 Å/s for the Ag layer.

Since the refractive index of a glass substrate (~1.52) is similar to that of $SiO_2$ (~1.46), the bottom-most $SiO_2$ layer of the DBSR was omitted in **Figure 1a**; instead, a $Ta_2O_5$ layer is present as the bottom-most layer.

A mixture of PDMS and curing agent at a mass ratio of 8:2 was prepared and spin-coated at 500 rpm for 15 seconds. The PDMS-coated sample was then placed in a vacuum chamber for 15 minutes to remove air bubbles and form a dense film, followed by baking at 90 °C for 15 minutes.

The overall reflection and transmission spectra (300–2500 nm) were measured using a UV-visible spectrophotometer (Lambda 1050, PerkinElmer), and the emissivity spectra in the mid-IR region were measured using Fourier-transform infrared spectroscopy (Nicolet iN10MX, ThermoFisher).

*Optical calculations:* The overall calculated *R* and *T* spectra were obtained using the transfer matrix method. The refractive indices of $SiO_2$ and $Ta_2O_5$ were measured using an ellipsometer, while that of Ag was extracted from the literature[27]. These are shown in **Figure S5**.

*Outdoor experiment:* The air temperatures inside the polystyrene boxes were measured using four K-type thermometers (Omega Engineering), connected with a temperature data logger (Omega Engineering, RDXL4SD). The ambient temperature was measured using a temperature sensor (Omega Engineering, RTD-805), which was placed under polystyrene foam with a hole, covered with aluminum foil, and connected to an external data logger (Omega Engineering, OM-CP-RTDTEMP101A). Solar irradiance was measured using a commercial photodiode (Thorlabs, S120VC) and calibrated using a solar simulator at 547.3 $W/m^2$.

## Supporting Information

PDF file available online

## Acknowledgements

At KAIST, this work was supported by the Creative Materials Discovery Program through the National Research Foundation of Korea (NRF) funded by the Ministry of Science and ICT (NRF-2016M3D1A1900035). At UW-Madison, work was supported primarily by internal funding.

## Author Contribution

Y. Jin and S. Kim contributed equally to this work. Y. Jin and K. Yu conceived the concept of the work. Y. Jin designed the optical structures and performed the optical calculations. S. Kim fabricated the samples and conducted the overall experiments, with contributions from Y. Jin. Y. Jin and M. Kats performed the cooling power analysis using a gradient atmospheric model.

Y. Jin and M. Kats wrote most of the paper, with contributions from all authors. K. Yu and M. Kats oversaw the research.

## Conflict of Interest

The authors declare no conflict of interest.

# Supporting Information

## Cool windows: simultaneously engineering high visible transparency and strong solar rejection

*Yeonghoon Jin,*[1,2,#] *Seungwon Kim,*[1,#] *Tanuj Kumar,*[2] *Mikhail A. Kats,*[2*] *and Kyoungsik Yu*[1,*]

[1]School of Electrical Engineering, Korea Advanced Institute of Science and Technology (KAIST), Daejeon 34141, Republic of Korea

[2]Department of Electrical and Computer Engineering, University of Wisconsin-Madison, Madison, WI 53706, USA

[#]These authors contributed equally

E-mail: mkats@wisc.edu, ksyu@kaist.edu

## Note S1. Distributed Bragg reflectors (DBRs) and our dual-band selective reflectors (DBSRs)

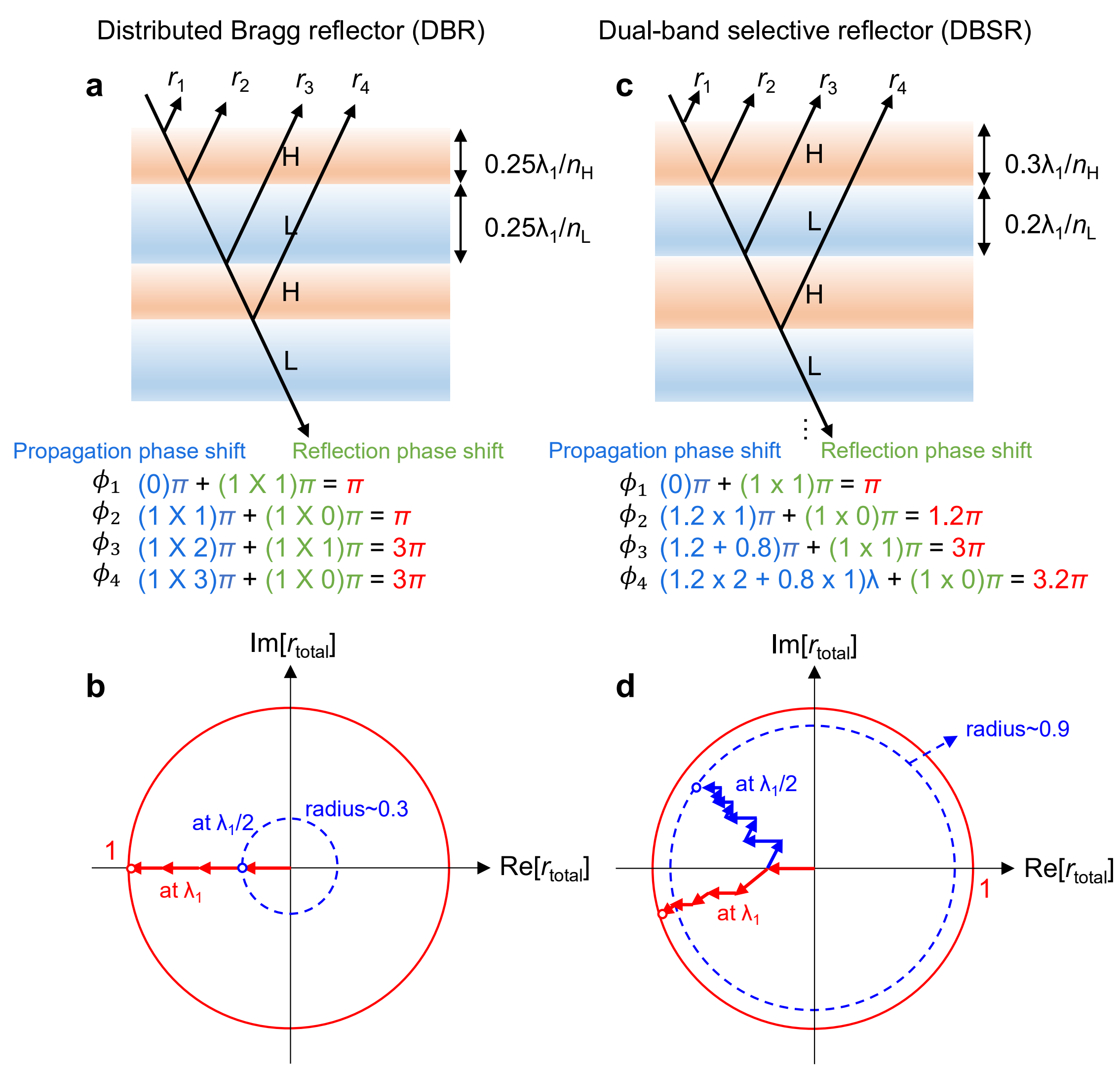

**Figure S1.** Working principles of a conventional distributed Bragg reflector (DBR) and our dual-band selective reflector (DBSR). **(a,c)** Depiction of partial reflected waves of **(a)** a DBR and **(c)** a DBSR, neglecting multiple reflections within the films. The total accumulated phase comprises the round-trip propagation phase and reflection phase. **(b,d)** Visual sum of the partial reflected waves of **(b)** a 5-pair DBR and **(d)** a 5-pair DBSR, both at $\lambda_1 = 780$ nm. In both cases, $SiO_2$ ($n_L$ = 1.46) is used as the low-index material and $Ta_2O_5$ ($n_L$ = 2.08) as the high-index material.

This section introduces the concept of a dua-band selective reflector (DBSR), which is a modified version of a distributed Bragg reflector (DBR), but where we can independently control the wavelengths of multiple reflection peaks. For this paper, we aim to simultaneously design reflectance peaks in the near-IR and near-UV.

DBRs create constructive interference of partially reflected waves (**Figures S1a,b**, though note that in this figure we only draw the primary partial reflections, neglecting multiple reflections within each film; this is a reasonable assumption assuming that the Fresnel reflection coefficients at each interface are not too high).[1]

**Figure S1a** shows the reflection phase shift ($\phi$) of the partial reflected waves (from $r_1$ to $r_4$) in a DBR, and **Figure S1b** shows the sum of the complex-reflectance partial waves of a DBR composed of 5 pairs of $SiO_2/Ta_2O_5$. Here, the target wavelength ($\lambda_{target}$) for the center of the photonic bandgap is 780 nm, and $SiO_2$ is used as the low-index material ($L$) and $Ta_2O_5$ as the high-index material ($H$); the thickness of $Ta_2O_5$ is 94 nm and that of $SiO_2$ is 134 nm, which corresponds to $\lambda_{target}/4n$, where $n$ is the refractive index. The amplitude of the sum of the overall reflected waves ($r_{total}$) approaches 1, and its phase is π.

To achieve total constructive interference, the reflected waves must be in phase. The reflection at the top interface $\phi_1$ (air to high index) introduces a phase shift of π, and thus, the phase of subsequent reflected waves ($\phi_2$, $\phi_3$, $\phi_4$…) must have phase of π, 3π, 5π, 7π, etc.

For example, the reflection phase shift of the second interface, $\phi_2$, is 0 (high to low index), and therefore, the round-trip propagation phase shift within the topmost layer must be an odd multiple of π. For the phase shift at the next interface $\phi_3$, the reflection phase shift (from low to high index) is π, so the round-trip propagation phase shift (in the first and second layers) should be an even multiple of π. Thus, the required round-trip propagation phase alternates between odd and even multiple of π (see the phase calculations in **Figure S1a**). To satisfy this, the round-trip propagation phase shift in a single layer must be an odd multiple of π. This condition can be written as:

$$2knd = (2m - 1)\pi, \tag{S1}$$

where $k$ is the wavenumber ($k = 2\pi/\lambda_m$), $n$ is the refractive index, $d$ is the thickness, and $m$ is a positive integer.[1] For the first mode ($m = 1$), the thickness $d$ should be $\lambda_1/4n$. Provided that the refractive index of $Ta_2O_5$ is 2.08 and that of $SiO_2$ is 1.45, the corresponding thickness of $Ta_2O_5$ is ~94 nm and that of $SiO_2$ is ~134 nm. When $n$ and $d$ are fixed, **Equation S1** is satisfied when $\lambda_m = 4nd/(2m - 1)$. For example, when $\lambda_1 = 4nd = 780$ nm, $\lambda_2 = 4nd/3 = \lambda_1/3 = 260$ nm and $\lambda_3 = 4nd/4 = \lambda_1/4 = 156$ nm. Accordingly, realizing high $R$ in both the 300–400 nm and 680–950 nm ranges is not possible this structure; the reflection amplitude is ~1 at $\lambda_1 = 780$ nm, whereas that at $\lambda_1/2 = 390$ nm is ~0.3 (**Figure S1b**).

Here, we design a structure to adjust wavelengths of high-order modes ($m$ = 2, 3…), while maintaining the wavelength and magnitude of the peak reflectance for the first mode. The idea is to slightly change the thickness from each layer from the $0.25\lambda_1/n$ condition, while having the round-trip phase shift in one pair remain 2π (**Figure S1c**). For example, let us assume that the thickness of the high-index layer is $0.3\lambda_1/n_H$ ($d_H$ = 112 nm, where $\lambda_1$ = 780 nm) and that of the low-index layer is $0.2\lambda_1/n_L$ ($d_L$ = 108 nm). Then, the round-trip propagation phase shift of the high-index layer is 1.2π ($2\times0.3\lambda_1/n_H$) and that of the low-index is 0.8π ($2\times0.2\lambda_1/n_L$), and then the accumulated phases of the odd-numbered reflected waves ($\phi_1$ = π, $\phi_3$ = 3π, and $\phi_5$ = 5π) remain the same as in the DBR case (compare **Figure S1a** and **Figure S1c**), while the accumulated phase of the even-numbered reflections are slightly off (by 0.2π) from ideal constructive interference. Nevertheless, the total reflectance $|r_{total}|$ at $\lambda_1$ remains close to 1 (**Figure S1d**). The phase condition of this structure can be expressed as $2kn_Hd_H + 2kn_Ld_L = 2m\pi$, which simplifies to $n_Hd_H + n_Ld_L = 0.5m\lambda_m$ ($n_Hd_H \neq n_Ld_L \neq 0.25m\lambda_m$). Given that the left term is a constant, the higher-order modes appear at $\lambda_2$ = 780/2 = 390 nm ($m$ = 2) and $\lambda_3$ = 780/3 = 260 nm. This indicates that high reflectance can be achieved around 390 nm (ultraviolet) and 780 nm (near-infrared). See the reflected wave trajectories at $\lambda_1$ =780 nm and $\lambda_1/2$ = 390 nm. in **Figure S1d**, which shows high reflection amplitude for both wavelengths. See **Figure 2c** in the main text for the reflectance spectra of the 5-pair DBSR.

The peak wavelength slightly shifts as the number of pair increases because of imperfect constructive interference, but the $R$ values in the 350–400 and 680–950 nm wavelength ranges remain high. Note that if the round-trip propagation phase shift in each layer ($2kn_Hd_H$ and $2kn_Ld_L$) deviates too far from π, the imperfect constructive interference will significantly reduce the reflectance (**Figure S2**).

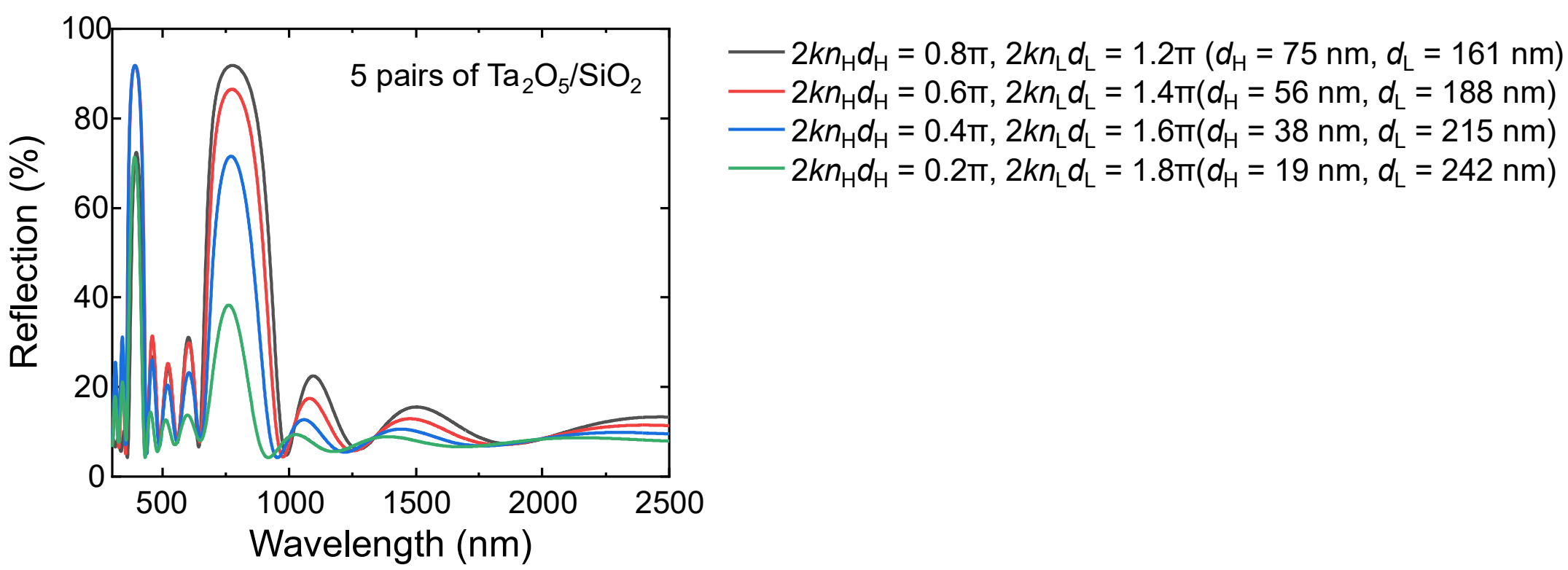

**Figure S2.** Reflection spectra of DBSRs with various thickness combinations (5 pairs).

In **Figure S3**, we compared the reflectance of the 6-pair DBR ($Ta_2O_5$ 94 nm/$SiO_2$ 134 nm) and the 6-pair DBSR ($Ta_2O_5$ 112 nm/$SiO_2$ 108 nm). Unlike the DBR, the DBSR shows high *R* in the 350–400 nm range, while there is almost no difference between the two in the 680–950 nm range.

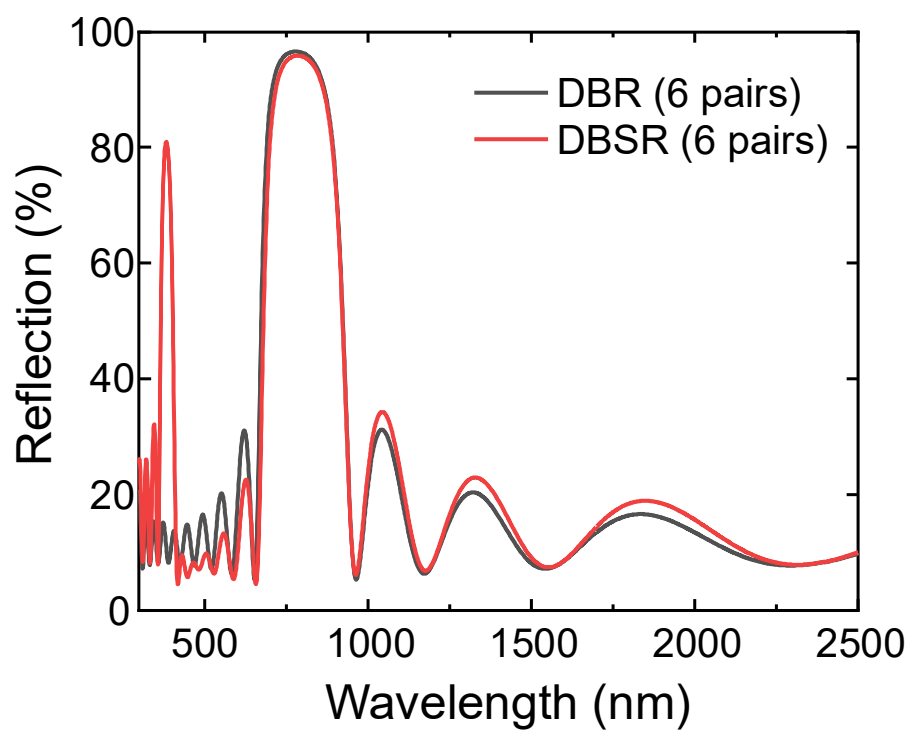

**Figure S3.** The reflection spectrum of the 6-pair DBR and the 6-pair DBSR.

## Note S2. Optical properties of samples and refractive indices of materials

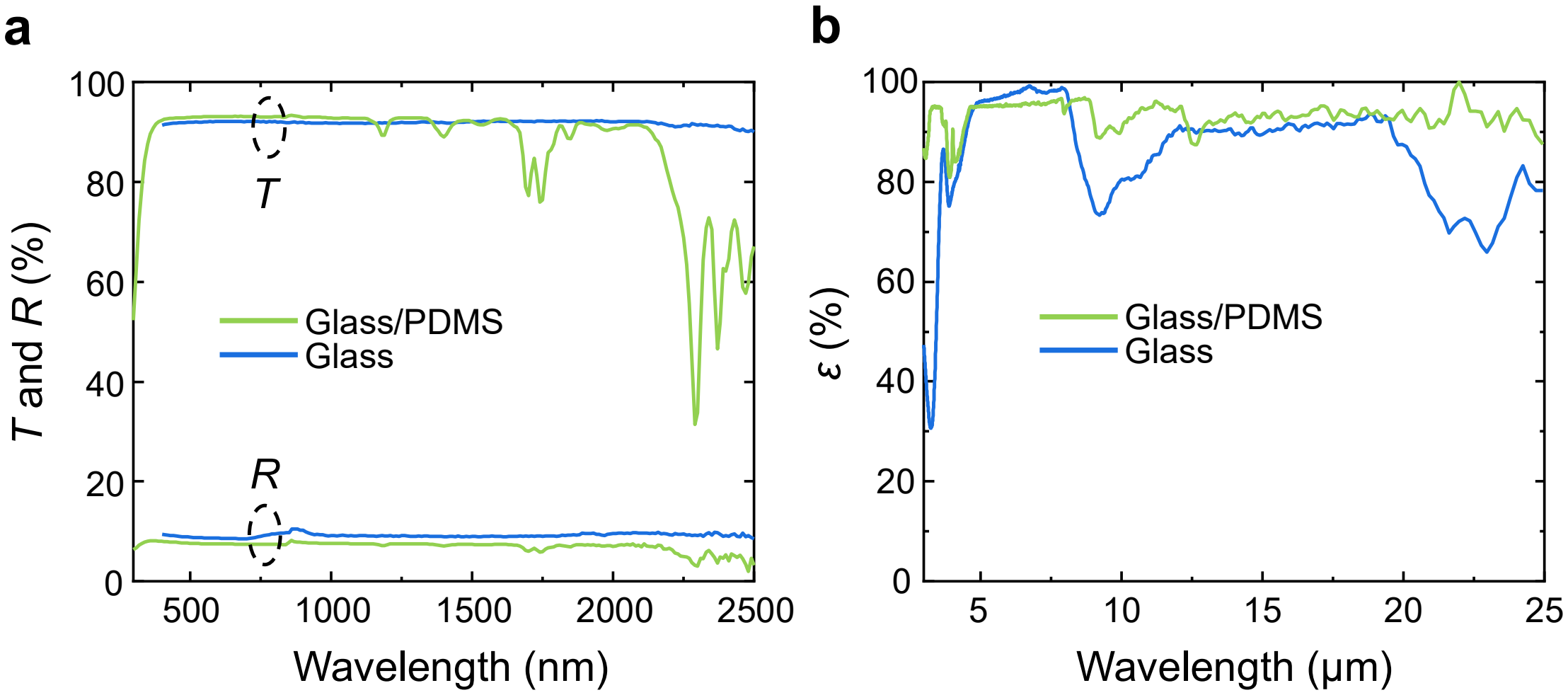

**Figure S4.** Optical properties of the glass and glass/PDMS (300 μm) samples. **(a)** Measured UV, visible, and near-IR transmittance (*T*) and reflectance (*R*), and **(b)** mid-infrared emissivity (*ε*) of both samples.

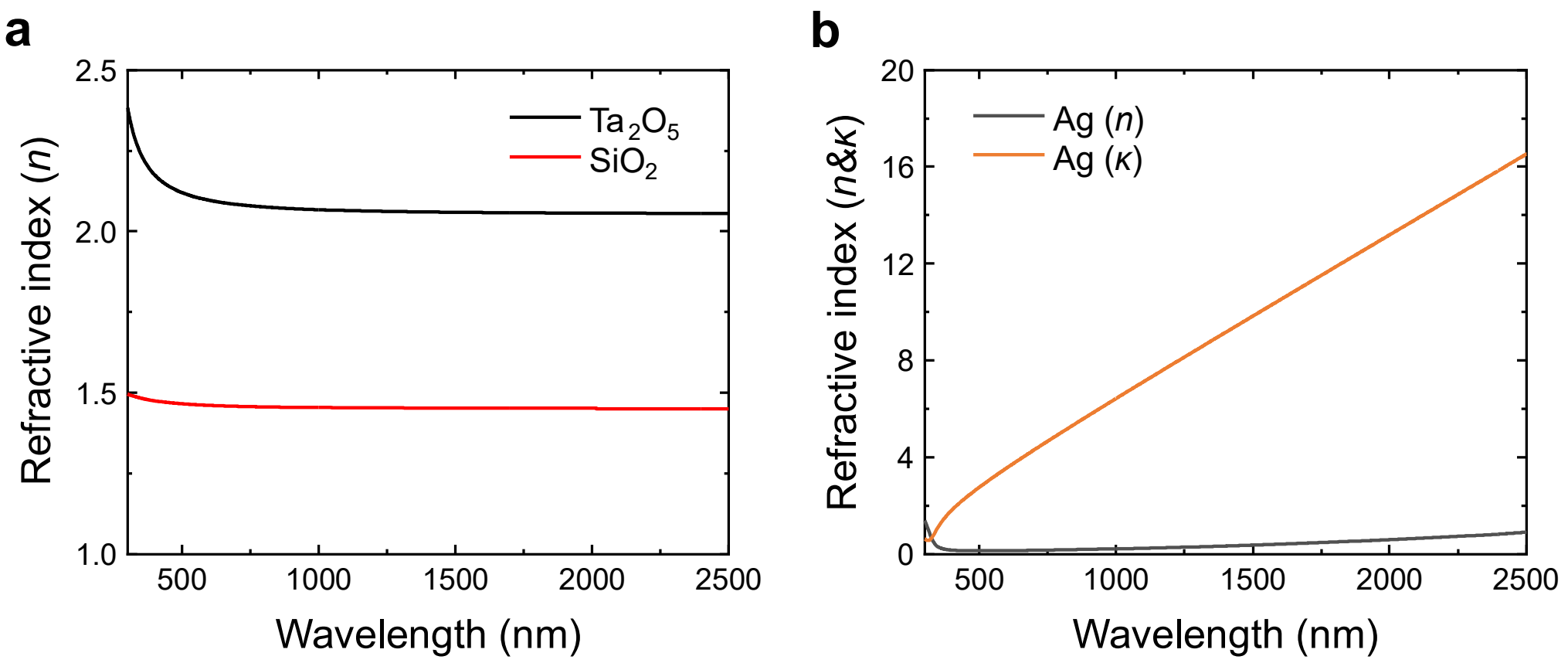

**Figure S5. (a)** The refractive indices of $Ta_2O_5$ and $SiO_2$ that were measured by an ellipsometer. **(b)** The complex refractive index of Ag that was extracted from the literature.[2]

## Note S3. A gradient atmospheric model to calculate the cooling performance

We recently demonstrated quantitatively that radiative cooling performance is significantly affected by altitude-dependent atmospheric gas composition and temperature, as well as by location and date/time.[3] However, most published papers have analysis that assumes a uniform atmosphere, where atmospheric gas composition and temperature are constant with respect to altitude. In reality, these parameters vary with altitude, and we found that the uniform atmospheric model typically underestimates cooling power by 10–40% compared to a gradient atmospheric model.[3] Here we summarize the relevant cooling calculations, the result of which is in **Figure 4(d-f)** in the main text.

Within the gradient atmospheric model, the net cooling power ($P_{\text{cool}}$) can be written as:[3]

$$P_{\text{cool}} = P_{\text{rad}}(T_{\text{surface}}) - \sum_{n=1}^{n_{atm}} P_{\text{atm}}^{n}(T_{\text{atm}}^{n}) - P_{\text{sun}} - P_{\text{con}}\left(T_{\text{surface}}, T_{\text{atm},1}\right). \quad \text{(S2)}$$

$P_{\text{rad}}$ is the power radiated from a cooling surface (here, the surface of our cool window), and can be written as:

$$P_{\text{rad}}(T_{\text{surface}}) = \int d\Omega \cos\theta \int_0^{\infty} d\lambda I_{\text{BB}}(T_{\text{surface}}, \lambda)\varepsilon_{\text{surface}}(\lambda, \theta), \quad \text{(S3)}$$

where $T_{\text{surface}}$ is the surface temperature, $\int d\Omega = 2\pi \int_0^{\pi/2} d\theta \sin\theta$, $\theta$ is the emission angle, $I_{\text{BB}}$ is the spectral radiance of a blackbody, and $\varepsilon_{\text{surface}}$ is the thermal emissivity of the surface. Here, the integration wavelength range is from 3 to 25 μm.

The second term represents the power radiated from the atmosphere and then absorbed by the surface (here, the window). Since atmospheric gas composition and temperature vary with altitude, our gradient model accounts for this by dividing the atmosphere into 10 layers; the partial atmospheric emissivity ($\varepsilon_{\text{atm}}$) from each atmospheric layer is shown in **Figure S6c**, with the temperature of each layer detailed in the figure caption. $P_{\text{atm}}^{n}$ in the second term represents the power radiated from the $n^{\text{th}}$ atmospheric layer and then absorbed by the cooler. $P_{\text{atm}}^{1}$ can be given by:[3]

$$P_{\text{atm}}^{1}(T_{\text{atm}}^{1}) = \int d\Omega \cos\theta \int_0^{\infty} d\lambda I_{\text{BB}}(T_{\text{atm}}^{1}, \lambda)\varepsilon_{\text{surface}}(\lambda, \theta)\varepsilon_{\text{atm}}^{1}(\lambda, \theta), \quad \text{(S4)}$$

where $T_{\text{atm}}^{1}$ is the temperature of the first atmospheric layer (290 K at 0–0.5 km), and $\varepsilon_{\text{atm}}^{1}$ is the emissivity of the first atmospheric layer. $P_{\text{atm}}^{n}$ can be given by ($n > 1$):[3]

$$P_{\text{atm}}^{n}(T_{\text{atm}}^{n}) = \int d\Omega \cos\theta \int_0^{\infty} d\lambda I_{\text{BB}}(T_{\text{atm}}^{n}, \lambda)\varepsilon_{\text{surface}}(\lambda, \theta)[\varepsilon_{\text{atm}}^{n}(\lambda, \theta) - \varepsilon_{\text{atm}}^{n-1}(\lambda, \theta)], \quad \text{(S5)}$$

and the atmospheric temperatures of each layer ($T_{\text{atm}}^{n}$) is shown in the caption of **Figure S6**.

The solar absorption can be given by: $P_{\mathrm{sun}} = \int_0^\infty d\lambda [1 - R_{\mathrm{surface}}(\lambda, \theta_{\mathrm{sun}})] I_{\mathrm{Solar}}(\lambda)$, where $\theta_{\mathrm{sun}}$ is the polar angle of the sun (in our case $\theta_{\mathrm{sun}} = 0$, where the sun is directly shining on the window surface), and $I_{\mathrm{Solar}}$ is the AM1.5 solar spectrum.[4] We assumed that, in our experiment, the incident solar energy to the polystyrene boxes is entirely absorbed by the black-painted boxed. The non-radiative heat exchange via conduction and convection can be given by: $P_{\mathrm{con}} = h_{\mathrm{c}}(T_{\mathrm{atm}}^1 - T_{\mathrm{cooler}})$, where $h_{\mathrm{c}}$ is a heat transfer coefficient.

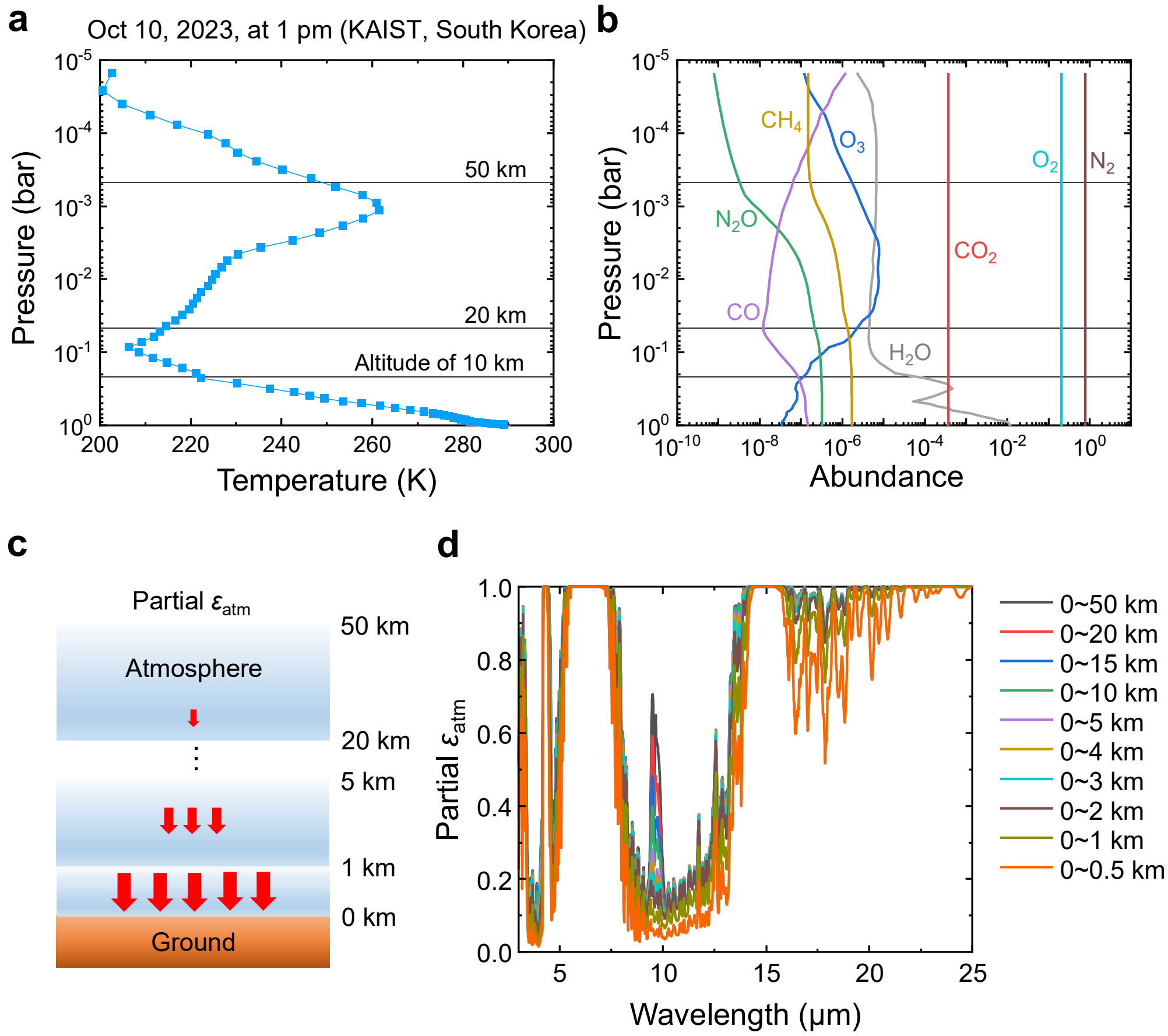

**Figure S6.** Altitude-dependent atmospheric information on October 10, 2023, at 1 pm at KAIST (36.37°N, 127.36°E). **(a)** Altitude-dependent atmospheric temperature profile. The y-axis represents both altitude and pressure. **(b)** Altitude-dependent atmospheric gas composition. The abundance indicates the ratio of the pressure of each gas to the total pressure at a given altitude. **(c)** Illustration of partial atmospheric emissivity ($\varepsilon_{\mathrm{atm}}$). **(d)** Partial atmospheric emissivity ($\varepsilon_{\mathrm{atm}}$) from 10 atmospheric layers. Most atmospheric radiation occurs from the first few kilometers above the ground, as the atmospheric gases are concentrated in this range. The atmospheric temperatures for each atmospheric layer, in order from the lowest altitude ($T_{\mathrm{atm,1}}$) to the highest altitude ($T_{\mathrm{atm,10}}$), are: 290, 286.3, 280.7, 275.9, 270, 262, 253.6, 220, 208, and 214 K.

**Figure S6a** shows the altitude-dependent atmospheric temperature profile at KAIST, South Korea (36.37°N, 127.36°E) on October 10$^{th}$, 2023, at 1 pm; this is the time of the experiment in **Figure 4**. The atmospheric data was obtained through NASA's Planetary Spectrum Generator (PSG);[5] refer to our recent paper[3] for instructions on how to use the PSG. **Figure S6b** shows the altitude-dependent atmospheric gas abundance on the same date/time. The gas abundance (*x*-axis) indicates the ratio of the pressure of each gas to the total gas pressure at a given altitude; for example, the abundance of $CO_2$, $O_2$, and $N_2$ is maintained across the entire altitude, but their pressure indeed decreases with increasing altitude because the total pressure (*y*-axis) decreases. Note that $H_2O$ and $CO_2$ are concentrated in the first few kilometers from the ground, indicating that most thermal emission from the atmosphere occurs within a few kilometers from the ground.

## Note S4. Energy-saving benefit of the cool window

Here we estimate the potential energy-saving benefit of our cool window relative to bare glass using a simplified enclosure model. We consider the same configuration used in our outdoor experiment, where the window is mounted on top of a polystyrene enclosure with black-painted interior surfaces, and all light transmitted through the window is assumed to be fully absorbed by the objects inside the enclosure.

Assuming an incident solar irradiance of 1000 W/m$^2$, the effective solar heat gain inside the enclosure is 484 W/m$^2$ for our cool window and 910 W/m$^2$ for bare glass. Therefore, the reduction in cooling load is 426 W/m$^2$. Because the emissivities of our cool window and bare glass are comparable, and because solar management is the dominant factor under daytime conditions, it is reasonable to treat this value as the primary difference in heat load between the two cases.

To convert this reduced thermal load into electrical energy savings for an air-conditioning system, we use the seasonal energy efficiency ratio (SEER), defined as:

$$\mathrm{SEER} = \frac{\text{the cooling power during a typical cooling season (BTU)}}{\text{the total electricity input (Wh)}}, \tag{S6}$$

where BTU is British thermal unit. SEER represents the cooling efficiency of an AC. The minimum SEER requirements for residential systems set by the US Department of Energy (DOE) is 14 BTU/Wh in 2023.

Provided that 1 W = 3.4 BTU/h, 426 W/m$^2$ equals to 1448 BTU/h/m$^2$. The corresponding electrical power saving is therefore:

$$\text{Electrical input saved} = \frac{1448\ \text{BTU/h/m}^2}{14\ \text{BTU/Wh}} = 103\ \text{W/m}^2. \tag{S7}$$

If we further assume that peak solar conditions requiring an AC persist for 5 h/day, the daily electricity saving becomes 515 Wh/m$^2$/day, and the annual saving becomes higher than 55 kWh/m$^2$/year. These values are first-order estimates intended to illustrate the potential practical impact of the proposed cool window.

### Note S5. A design to prevent the oxidation and sulfidation of a thin Ag layer

Our multilayer structure is composed of $SiO_2$, $Ta_2O_5$, and a thin Ag layer. The oxides are widely used as environmentally stable optical coating materials with good resistance to oxidation, moisture, and UV exposure. The primary concern is therefore the long-term stability of Ag.

We note that thin Ag films have been extensively used in practical optical coatings because of their low optical loss, and multiple protection strategies have been demonstrated in the literature. For example, a 15-nm-thick alumina ($Al_2O_3$) protection film deposited by atomic layer deposition (ALD) on top of an Ag layer preserved the optical performance of the Ag film even under harsh oxygen-plasma conditions.[6] In addition, thin oxide films on top of an Ag layer showed minimal degradation after 16 months under humid environments (relative humidity of 60~85%).[7]

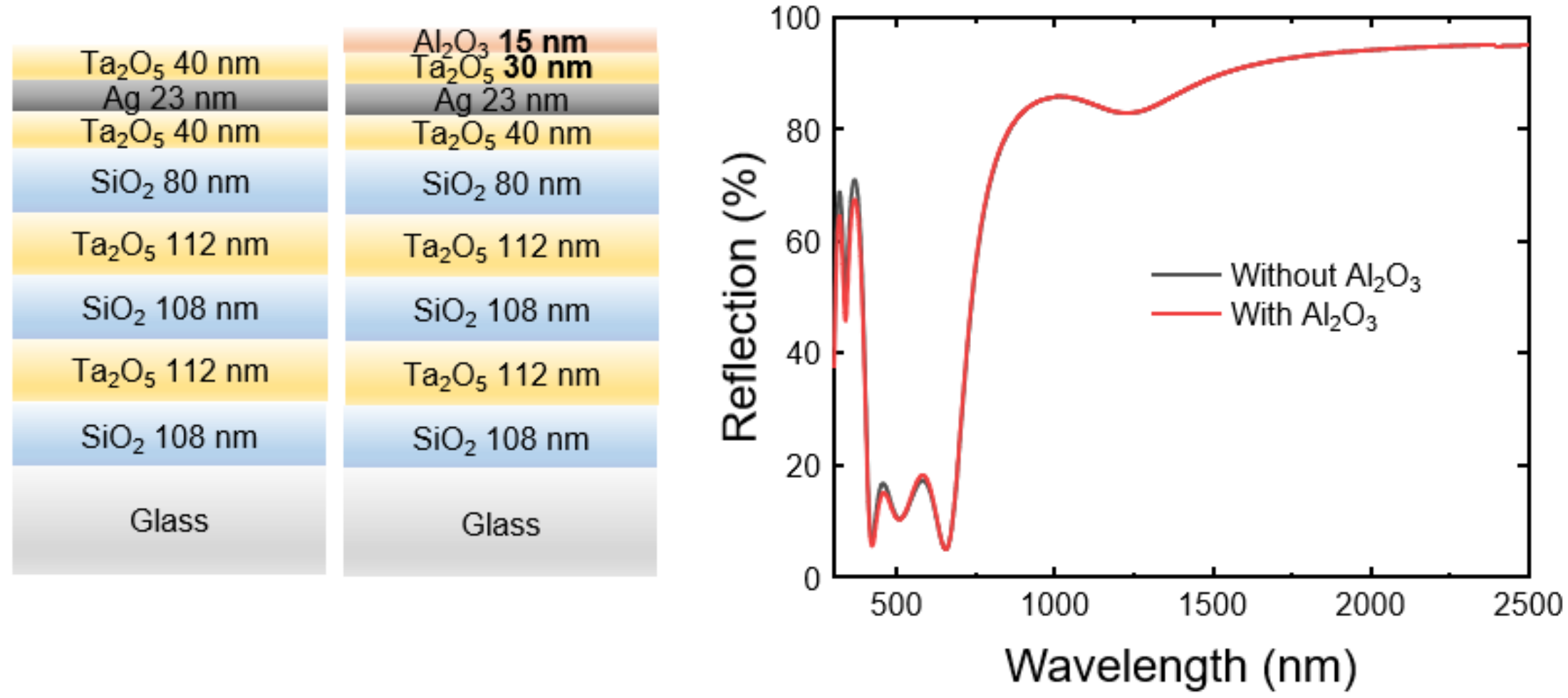

**Figure S7.** Reflectance spectra of our cool window with and without an encapsulation $Al_2O_3$ layer.

These strategies are compatible with our structure. For example, we can deposit 15-nm-thick $Al_2O_3$ covering our photonic structure as additional protection. As shown in **Figure S7**, the reflection spectra with and without thin $Al_2O_3$ layer are almost identical.